\documentclass[titlepage,12pt]{article}
\usepackage{amssymb}
\usepackage{amsfonts}
\usepackage{geometry}
\usepackage{mathrsfs}
\usepackage[onehalfspacing]{setspace}

\input{tcilatex}
\renewcommand{\mathcal}{\mathscr}

\begin{document}

\title{Exploring patterns of demand in bike sharing systems via replicated
point process models}
\author{Daniel Gervini and Manoj Khanal \\
%EndAName
Department of Mathematical Sciences\\
University of Wisconsin--Milwaukee}
\maketitle

\begin{abstract}
Understanding patterns of demand is fundamental for fleet management of bike
sharing systems. In this paper we analyze data from the Divvy system of the
city of Chicago. We show that the demand of bicycles can be modeled as a
multivariate temporal point process, with each dimension corresponding to a
bike station in the network. The availability of daily replications of the
process allows nonparametric estimation of the intensity functions, even for
stations with low daily counts, and straightforward estimation of pairwise
correlations between stations. These correlations are then used for
clustering, revealing different patterns of bike usage.

\end{abstract}

\section{Introduction\label{sec:Intro}}

Bike sharing systems are becoming increasingly common in large cities around
the world (Shaheen et al., 2010). These systems provide short-term bicycle
rental services at unattended stations distributed throughout the city. A
user checks out a bicycle at a station near the intended origin of the
journey and returns it at a station near the intended destination. For the
system to run smoothly, it is necessary that both bicycles and docks be
available at every station. When no bicycles are available at the intended
origin of a trip or no docks are available at the intended destination,
users needs to look for alternative nearby stations, which may dissuade them
from using the system altogether. Since bike flow from one station to
another is rarely matched by a similar flow in the reverse direction,
imbalances in the spatial distribution of bikes inevitably arise (Nair and
Miller-Hooks, 2011). There are different strategies to manage this problem.
For example, bikes are manually relocated by trucks as part of the
day-to-day operations of the system. From a longer-term perspective, careful
planning of the location of new stations is important. In order to make good
short- and long-term decisions, understanding the spatiotemporal patterns of
bike demand is fundamental.

In this paper we show that bike demand at each station can be modeled as a
temporal point process, where bike checkout times are the random events of
interest. Bike return can be modeled in a similar way. We will analyze data
from the Divvy system of the city of Chicago, publicly available at the
Chicago Data Portal website (https://data.cityofchicago.org). Specifically,
we will analyze bike trips that took place between April 1 and November 30
of 2016, since bike usage considerably decreases during the winter. There
were a total of $3,068,211$ bike trips and 458 active bike stations in that
period. Demand varied a lot depending on the station location, from a lowest
of 29 annual trips for station 386 in the South Side to a highest of $85,314$
annual trips for station 35 at the Navy Pier. For stations with relatively
large daily counts, the distribution of bike demand on any given day can be
estimated by kernel smoothing or other density estimation methods
(Silverman, 1986). But for stations with low daily counts this is not
possible, at least not in a meaningful way. In this paper we propose a new
method that overcomes this deficiency by \textquotedblleft
borrowing\textquotedblright\ data across replications, i.e.~across different
days. In the end, estimators of daily distributions of bike demand are
obtained, even for stations with low daily counts, but in an indirect way.
These estimators are then used to study spatial correlations between
stations and to derive clusters that correspond to different usage patterns.

To avoid confusion, let us clarify that in this paper we use the terms
`realization', `replication' and \ `observation' the way they are used in
functional data analysis, which is somewhat different from the point-process
literature. By `realization' or `replication', which for us are synonyms, we
refer to a realization of the whole process, that is, the whole set of
observations on any given day. By `observation' we refer to an individual
point in a realization of the process, that is, to a bike trip on any given
day. Thus, our data set contains 244 replications or realizations of the
process, each with a varying number of observations.

To put the problem in context, we note that different aspects of bike
sharing systems have been studied in the specialized literature
(e.g.~Borgnat \emph{et al.}, 2011; Vogel \emph{et al.}, 2011; Nair \emph{et
al.}, 2013), but the problem of estimating and modeling daily demand
distribution at every station in a network has not yet been addressed, to
the best of our knowledge. From a statistical methodology perspective, we
can mention early work on replicated point processes by Diggle \emph{et al.}%
~(1991), Baddeley \emph{et al.}~(1993), Diggle \emph{et al.}~(2000), Mateu
(2001) and Landau \emph{et al.~}(2004), but these authors propose tests for
various hypotheses using summary statistics of the process (see also
Baddeley \emph{et al.}, 2015, ch.~16; Diggle, 2013, ch.~5.4), rather than
explicitly estimating the intensity functions of the processes, as we do
here. More recent work that does address the intensity-function estimation
problem was done by Wu \emph{et al.}~(2013), Bouzas and Ruiz-Fuentes (2015)
and Gervini (2016), but only in the context of univariate processes, not
multivariate ones as in this paper. Spatio-temporal processes have been
widely studied in the literature, but mostly in the single-replication
context (see e.g.~Li and Guan, 2014; Shirota and Gelfand, 2017; Diggle,
2013, and references therein), not in the many-replication context of this
paper. Finally, we mention that clustering methods for spatial functional
data have been proposed by Delicado \emph{et al.}~(2010), Romano \emph{et al.%
}~(2010), Secchi \emph{et al.}~(2013), and Menafoglio and Secchi (2017),
among others, but again in the context of a single datum per site, which
does not allow direct estimation of spatial correlations and requires
assumptions such as isotropy; in our application, the availability of many
replications per site allows us to estimate spatial correlations directly
and without isotropy assumptions, which, in fact, we show not to hold for
the bike sharing network.

\section{Modeling daily bike demand\label{sec:Model}}

\subsection{Poisson point processes}

Let $X_{ij}$ be the set of checkout times for day $i$ at bike station $j$.
In our data set we have $n=244$ days and $d=458$ bike stations. Each $X_{ij}$
is a finite but otherwise random set, so it is best modeled as a point
process. The collection $\mathbf{X}_{i}=(X_{i1},\ldots ,X_{id})$ can be seen
as a realization of a multivariate point process. For an overview of point
processes, see M\o ller and Waagepetersen (2004, ch.~2), Streit (2010,
ch.~2) or Baddeley (2007).

A temporal point process $X$ is a random countable set in $\mathbb{[}%
0,\infty )$. A process is locally finite if $\#(X\cap B)<\infty $ with
probability one for any bounded interval $B$, in which case we can define
the count function $N(B)=\#(X\cap B)$. A Poisson process is a locally finite
process for which there exists a nonnegative locally integrable function $%
\lambda (t)$ such that \emph{(i)} $N(B)$ follows a Poisson distribution with
rate $\int_{B}\lambda (t)dt$ for any bounded $B$, and \emph{(ii)} for
disjoint sets $B_{1},\ldots ,B_{k}$ the random variables $N(B_{1}),\ldots
,N(B_{k})$ are independent. A consequence of \emph{(i)} and \emph{(ii)} is
that the conditional distribution of the points in $X\cap B$ given $N(B)=m$
is the distribution of $m$ independent and identically distributed
observations with density $\lambda (t)/\int_{B}\lambda $. The function $%
\lambda $ is called the intensity function of the process.

\subsection{The model}

In our application we have $nd$ processes $X_{ij}$ with $i=1,\ldots ,n$ and $%
j=1,\ldots ,d$, each with a corresponding intensity function $\lambda _{ij}$
on the interval $[a,b]=[0,24]$. Since the $\lambda _{ij}$s are nonnegative,
for simplicity we will assume that they are positive everywhere, even if
negligible in some regions, and model their logarithms using additive
principal component models, similar to those used in functional data
analysis (Ramsay and Silverman, 2005, ch.~8).

For each station $j$ we assume 
\begin{equation}
\log \lambda _{ij}(t)=\mu _{j}(t)+\sum_{k=1}^{p_{j}}u_{ikj}\phi _{kj}(t),\ \
\ i=1,\ldots ,n,  \label{eq:log_lmb_model}
\end{equation}%
where $\mu _{j}(t)$ is the annual mean function for station $j$ and $\{\phi
_{kj}\}_{k=1}^{p_{j}}$ are orthonormal functions (across $k$s) that account
for various types of deviations from the mean. We will refer to the $\phi
_{kj}$s as components and the $u_{ikj}$s as component scores. The component
scores are, in principle, random effects with $E(u_{ikj})=0$, $\sigma
_{kj}^{2}=V(u_{ikj})$ and $\limfunc{cov}(u_{ikj},u_{ik^{\prime }j})=0$ for $%
k\neq k^{\prime }$. Without loss of generality we assume $\sigma
_{1j}^{2}\geq \cdots \geq \sigma _{p_{j}j}^{2}>0$. However, for estimation
purposes we will treat the $u_{ikj}$s as fixed effects, which does not
require distributional assumptions on the $u_{ikj}$s that may be
questionable.

Model (\ref{eq:log_lmb_model}) for $\log \lambda _{ij}$ turns into a
multiplicative model for $\lambda _{ij}$: 
\begin{equation}
\lambda _{ij}(t)=\lambda _{0j}(t)\prod_{k=1}^{p_{j}}\psi _{kj}(t)^{u_{ijk}},
\label{eq:lmb_model}
\end{equation}%
where $\lambda _{0j}(t)=\exp \mu _{j}(t)$ and $\psi _{kj}(t)=\exp \phi
_{kj}(t)$. We will refer to $\lambda _{0j}(t)$ as the baseline intensity
function for station $j$.

Since the $\lambda _{ij}$s are not directly observable, the mean $\mu
_{j}(t) $ and the components $\phi _{kj}(t)$ must be estimated from the
data. To facilitate this, we use spline models (De Boor, 1978): 
\begin{equation}
\mu _{j}(t)=\sum_{l=1}^{q}c_{l0j}\beta _{l}(t)\text{,\ \ \ }\phi
_{kj}(t)=\sum_{l=1}^{q}c_{lkj}\beta _{l}(t),  \label{eq:spline_model}
\end{equation}%
where $\{\beta _{l}\}_{l=1}^{q}$ is a spline basis. We use B-splines in this
paper, but other bases can be used, even non-spline bases such as the
Fourier basis. Modeling $\mu _{j}$ and the $\phi _{kj}$s as spline functions
turns the functional estimation problem into a simpler multivariate problem
of estimating basis coefficients $\mathbf{c}_{kj}=(c_{1kj},\ldots
,c_{qkj})^{T}$. It also simplifies the introduction of periodicity
constraints: the intensity functions should satisfy $\lambda
_{ij}(0)=\lambda _{ij}(24)$ in this application, which is enforced by the
simple linear constraints $\mathbf{c}_{kj}^{T}\mathbf{\beta }(a)=\mathbf{c}%
_{kj}^{T}\mathbf{\beta }(b)$, where $\mathbf{\beta }(t)=(\beta
_{1}(t),\ldots ,\beta _{q}(t))^{T}$. Similarly, the orthonormality of the $%
\phi _{kj}$s is enforced by the constraints $\mathbf{c}_{kj}^{T}\mathbf{Jc}%
_{k^{\prime }j}=$ $\delta _{k,k^{\prime }}$, where $\mathbf{J}=\int_{a}^{b}%
\mathbf{\beta }(t)\mathbf{\beta }(t)^{T}dt$ and $\delta _{k,k^{\prime }}$ is
Kronecker's delta.

\subsection{Estimation}

Fitting model (\ref{eq:log_lmb_model}), then, involves estimation of the
parameters $\mathbf{c}_{kj}$s in (\ref{eq:spline_model}) and of the
component scores $u_{ikj}$, which, for estimation purposes, will be treated
as fixed effects. We do this by maximum likelihood, using the Poisson model
as working model. In view of the above-mentioned properties of the Poisson
process, the density function of $X_{ij}=\{t_{ij1},\ldots ,t_{ijm_{ij}}\}$
is 
\begin{equation}
f_{ij}(m_{ij},t_{ij1},\ldots ,t_{ijm_{ij}})=\exp \left\{
-\int_{a}^{b}\lambda _{ij}(t)dt\right\} \frac{1}{m_{ij}!}%
\prod_{l=1}^{m_{ij}}\lambda _{ij}(t_{ijl}),  \label{eq:pdf}
\end{equation}%
where $f_{ij}(0,\emptyset )=\exp \left( -\int_{a}^{b}\lambda _{ij}\right) $
if $X_{ij}=\emptyset $ and $m_{ij}=0$. Then the log-likelihood function for
station $j$, ignoring the constant factor $1/m_{ij}!$, is 
\begin{equation}
\ell _{j}=-\sum_{i=1}^{n}\int \lambda
_{ij}+\sum_{i=1}^{n}\sum_{l=1}^{m_{ij}}\log \lambda _{ij}(t_{ijl}).
\label{eq:Log_lik}
\end{equation}

In principle, the estimators $\mathbf{\hat{c}}_{kj}$s and $\hat{u}_{ikj}$s
would be the maximizers of $\ell _{j}$. However, maximizing $\ell _{j}$
without any sort of roughness penalty will produce irregular estimators of $%
\mu _{j}$ and the $\phi _{kj}$s if the spline basis dimension $q$ is large.
The roughness of a function $g$ can be measured by the functional norm of
its second derivative, $\int_{a}^{b}(g^{\prime \prime })^{2}$. So we will
define the $\mathbf{\hat{c}}_{kj}$s and preliminary estimators of the scores 
$\tilde{u}_{ikj}$s as the maximizers of the penalized log-likelihood
function 
\begin{eqnarray}
P\ell _{j} &=&\frac{1}{n}\ell _{j}-\xi _{1}\int (\mu _{j}^{\prime \prime
})^{2}-\xi _{2}\sum_{k=1}^{p_{j}}\int (\phi _{kj}^{\prime \prime })^{2}
\label{eq:Pen_log_lik} \\
&=&\frac{1}{n}\ell _{j}-\xi _{1}\mathbf{c}_{0j}^{T}\mathbf{\Omega c}%
_{0j}-\xi _{2}\sum_{k=1}^{p_{j}}\mathbf{c}_{kj}^{T}\mathbf{\Omega c}_{kj}, 
\nonumber
\end{eqnarray}%
where $\mathbf{\Omega }=\int_{a}^{b}\mathbf{\beta }^{\prime \prime }(t)%
\mathbf{\beta }^{\prime \prime T}(t)dt$ and $\xi _{1}$ and $\xi _{2}$ are
non-negative tuning parameters that regulate the degree of smoothness of $%
\mu _{j}$ and the $\phi _{kj}$s, respectively. The maximization has to be
carried out subject to the periodicity and orthonormality constraints 
\begin{eqnarray*}
\mathbf{c}_{kj}^{T}\mathbf{\beta }(a) &=&\mathbf{c}_{kj}^{T}\mathbf{\beta }%
(b),\ \ \ k=0,\ldots ,p_{j}, \\
\mathbf{c}_{kj}^{T}\mathbf{Jc}_{k^{\prime }j} &=&\delta _{k,k^{\prime }},\ \
\ k,k^{\prime }=1,\ldots ,p_{j},
\end{eqnarray*}%
and, since the true random effects $u_{ikj}$ are zero-mean uncorrelated
variables (across $k$s), we also impose the following constraints on the $%
u_{ikj}$s for estimation: 
\begin{eqnarray*}
\frac{1}{n}\sum_{i=1}^{n}u_{ikj} &=&0,\ \ \ k=1,\ldots ,p_{j}, \\
\frac{1}{n}\sum_{i=1}^{n}u_{ikj}u_{ik^{\prime }j} &=&0,\ \ \ k,k^{\prime
}=1,\ldots ,p_{j},\ \ \ k\neq k^{\prime }.
\end{eqnarray*}

Our preliminary simulations showed that the sample variances of the $\tilde{u%
}_{ikj}$s obtained this way tend to overestimate the true variances of the $%
u_{ikj}$s. To ameliorate this problem we re-scale the component scores,
letting $\hat{u}_{ikj}=\tau _{j}\tilde{u}_{ikj}$ and finding the optimal $%
\hat{\tau}_{j}$ by maximum likelihood based on the $m_{ij}$s. That is, since 
$m_{ij}\sim \mathcal{P}(\int_{a}^{b}\lambda _{ij}(t)dt)$ for a Poisson
process, the log-likelihood of the $m_{ij}$s is 
\[
\tilde{\ell}_{j}=-\sum_{i=1}^{n}I_{ij}(\tau )+\sum_{i=1}^{n}m_{ij}\log
I_{ij}(\tau ), 
\]%
where $I_{ij}(\tau )=\int_{a}^{b}\hat{\lambda}_{ij}^{(\tau )}(t)dt$ and $%
\hat{\lambda}_{ij}^{(\tau )}(t)$ is as in (\ref{eq:lambda_hat}) with $%
u_{ikj} $ replaced by $\tau \tilde{u}_{ikj}$. Then $\hat{\tau}_{j}$ is the
maximizer of $\tilde{\ell}_{j}$, and $\hat{u}_{ikj}=\hat{\tau}_{j}\tilde{u}%
_{ikj}$.

Fully-detailed algorithms to compute these estimators are explained in the
Supplementary Material, and Matlab programs are available on the first
author's website.

Once the mean $\mu _{j}$, the components $\phi _{kj}$s and the scores $%
u_{ikj}$s have been estimated, the individual daily intensity functions can
be estimated from model (\ref{eq:log_lmb_model}) as 
\begin{equation}
\hat{\lambda}_{ij}(t)=\exp \left\{ \hat{\mu}_{j}(t)+\sum_{k=1}^{p_{j}}\hat{u}%
_{ikj}\hat{\phi}_{kj}(t)\right\} .  \label{eq:lambda_hat}
\end{equation}%
They can subsequently be used for spatial inference regarding, for instance,
cross-correlations among bike stations, as we do in Section \ref%
{sec:Clustering}.

\subsection{Choice of tuning parameters}

The models introduced above have a number of tuning parameters that have to
be chosen by the user: the number of components $p_{j}$ in (\ref%
{eq:log_lmb_model}), the type and dimension $q$ of basis functions in (\ref%
{eq:spline_model}), and the smoothing parameters $\xi _{1}$ and $\xi _{2}$
in (\ref{eq:Pen_log_lik}). The specific type of basis functions does not
have much of an impact on the final estimator, provided the dimension is
large enough; we simply take cubic $B$-splines with equally spaced knots in
our simulations and data analyses in this paper. The dimension $q$ is more
relevant and should be relatively large, since the regularity of the
estimators will be taken care of by $\xi _{1}$ and $\xi _{2}$ (Eilers and
Marx, 1996); but for the same reason, it is not necessary to agonize over an
exact choice of $q$. As noted by Ruppert (2002, sec.~3), although $q$ can be
chosen systematically by cross-validation, there is little change in
goodness of fit after a minimum dimension $q$ has been reached, and the fit
is essentially determined by the smoothing parameters thereafter.

The choice of $\xi _{1}$ and $\xi _{2}$, then, is more important, and we do
it by cross-validation (Hastie \emph{et al.}, 2009, ch.~7). Leave-one-out
cross-validation finds $\hat{\xi}_{1j}$ and $\hat{\xi}_{2j}$ that maximize 
\[
\mathrm{CV}_{j}(\xi _{1},\xi _{2})=\sum_{i=1}^{n}\log \hat{f}%
_{ij}^{[-ij]}(m_{ij},t_{ij1},\ldots ,t_{ijm_{ij}}), 
\]%
where $\hat{f}_{ij}^{[-ij]}$ denotes the density (\ref{eq:pdf}) estimated
without observation $X_{ij}$. A faster alternative is to use $k$-fold
cross-validation, where the data is split into $k$ subsets that are
alternatively used as test data. We use five-fold cross-validation in our
implementation of the method.

The choice of the number of components $p_{j}$ can also be done by
cross-validation or, more practically, by the usual ad-hoc methods for
choosing the number of principal components (Jolliffe, 2002, ch.~6), which
take into account the relative contribution of the estimated variances $\hat{%
\sigma}_{kj}^{2}$ and stop at a $p_{j}$ where further additions of
components have a negligible impact on $\hat{\sigma}_{1j}^{2}+\cdots +\hat{%
\sigma}_{p_{j}j}^{2}$.

\section{Spatial correlations and clustering\label{sec:Clustering}}

\subsection{Measuring spatial correlation}

In multivariate analysis, a measure of overall correlation between two
random vectors $\mathbf{U}$ and $\mathbf{V}$ is the canonical correlation
coefficient $\rho =\max_{\mathbf{a},\mathbf{b}}\func{corr}(\mathbf{a}^{T}%
\mathbf{U},\mathbf{b}^{T}\mathbf{V})$, the largest possible correlation
between linear combinations of $\mathbf{U}$ and $\mathbf{V}$ (Izenman, 2008,
ch.~7.3). This coefficient can be computed as follows: given $\mathbf{\Sigma 
}_{UU}$ the covariance matrix of $\mathbf{U}$, $\mathbf{\Sigma }_{VV}$ the
covariance matrix of $\mathbf{V}$, and $\mathbf{\Sigma }_{UV}$ the
cross-covariance matrix of $\mathbf{U}$ and $\mathbf{V}$, then $\rho ^{2}$
is the largest eigenvalue of $\mathbf{\Sigma }_{UU}^{-1/2}\mathbf{\Sigma }%
_{UV}\mathbf{\Sigma }_{VV}^{-1}\mathbf{\Sigma }_{VU}\mathbf{\Sigma }%
_{UU}^{-1/2}$, or equivalently, of $\mathbf{\Sigma }_{VV}^{-1/2}\mathbf{%
\Sigma }_{VU}\mathbf{\Sigma }_{UU}^{-1}\mathbf{\Sigma }_{UV}\mathbf{\Sigma }%
_{VV}^{-1/2}$. The sample canonical correlation coefficient is obtained by
substituting sample covariance matrices.

In functional data analysis, where $U(t)$ and $V(t)$ are square-integrable
random functions, an equivalent version is defined (Horv\'{a}th and
Kokoszka, 2012): $\rho =\max_{\alpha ,\beta }\limfunc{corr}(\langle \alpha
,U\rangle ,\langle \beta ,V\rangle )$, where $\alpha $ and $\beta $ are
square-integrable functions and $\langle f,g\rangle =\int f(t)g(t)dt$. As we
show in the Appendix, the computation of $\rho $ can ultimately be reduced
to the multivariate case by using the principal component scores of $U(t)$
and $V(t)$, and the sample version is obtained by substituting sample
covariance functions and estimated component scores.

In our application, we are interested in the correlations of bike demand
between different stations, say $j$ and $j^{\prime }$, so we compute the
sample functional canonical correlation coefficient $\hat{\rho}_{jj^{\prime
}}$ of their respective log-intensity functions, 
\begin{equation}
\hat{\rho}_{jj^{\prime }}=\max_{\alpha ,\beta }\limfunc{corr}_{i=1,\ldots
,n}(\langle \alpha ,\log \hat{\lambda}_{ij}\rangle ,\langle \beta ,\log \hat{%
\lambda}_{ij^{\prime }}\rangle ).  \label{eq:rho_hat}
\end{equation}%
As explained above, $\hat{\rho}_{jj^{\prime }}^{2}$ is the largest
eigenvalue of $\mathbf{S}_{jj}^{-1/2}\mathbf{S}_{jj^{\prime }}\mathbf{S}%
_{j^{\prime }j^{\prime }}^{-1}\mathbf{S}_{j^{\prime }j}\mathbf{S}%
_{jj}^{-1/2} $, or equivalently of $\mathbf{S}_{j^{\prime }j^{\prime
}}^{-1/2}\mathbf{S}_{j^{\prime }j}\mathbf{S}_{jj}^{-1}\mathbf{S}_{jj^{\prime
}}\mathbf{S}_{j^{\prime }j^{\prime }}^{-1/2}$, where 
\[
\mathbf{S}_{jj^{\prime }}=\frac{1}{n}\sum_{i=1}^{n}(\mathbf{\hat{u}}_{ij}-%
\overline{\mathbf{\hat{u}}}_{\cdot j})(\mathbf{\hat{u}}_{ij^{\prime }}-%
\overline{\mathbf{\hat{u}}}_{\cdot j^{\prime }})^{T}, 
\]%
$\mathbf{\hat{u}}_{ij}=(\hat{u}_{i1j},\ldots ,\hat{u}_{ip_{j}j})^{T}$ and $%
\overline{\mathbf{\hat{u}}}_{\cdot j}=$ $\sum_{i=1}^{n}\mathbf{\hat{u}}%
_{ij}/n$.

\subsection{Spatial clustering}

Up to this point, we have treated the $d=458$ bike stations in our
application as generic dimensions of a multivariate point process, fitting
model (\ref{eq:log_lmb_model}) independently for each $j=1,\ldots ,d$.
However, when the $d$ dimensions correspond to $d$ locations in space, as in
this case, there is a spatial aspect to the problem that is interesting to
investigate.

We can think of the functional canonical correlation coefficients $\rho
_{jj^{\prime }}$ defined above as discretizations of a spatial correlation
function $R$, $\rho _{jj^{\prime }}=R(\mathbf{s}_{j},\mathbf{s}_{j^{\prime
}})$, where $\mathbf{s}_{j}$ and $\mathbf{s}_{j^{\prime }}$ are the spatial
coordinates of bike stations $j$ and $j^{\prime }$. In applications of
spatial functional data analysis where only one observation per site is
available (e.g.~Delicado et al., 2010; Menafoglio and Secchi, 2017),
estimation of $R$ requires assumptions such as isotropy, i.e.~that $R(%
\mathbf{s}_{j},\mathbf{s}_{j^{\prime }})=g(\Vert \mathbf{s}_{j}-\mathbf{s}%
_{j^{\prime }}\Vert )$ for some $g$, in order to pool data across
neighboring sites. But in our case, the availability of $n$ replications per
site allow us straightforward estimation of $\rho _{jj^{\prime }}$ by (\ref%
{eq:rho_hat}) without any assumptions on $R$. In fact, we will show in
Section \ref{sec:Example} that isotropy does not hold for our data.

The correlations $\hat{\rho}_{jj^{\prime }}$ can be used, for instance, to
discover clusters among bike stations. They can be obtained by applying
standard agglomerative techniques (Izenman 2008, ch.~12.3; Hastie \emph{et
al.}, 2009, ch.~14.3.12) to distances defined by $d_{jj^{\prime }}=1-\hat{%
\rho}_{jj^{\prime }}$. For our application we found that complete linkage
generally produces better results than either single or average linkage.

When the dimension $d$ is large, the number of different pairs $(j,j^{\prime
})$ can be extremely large; for example, $d(d-1)/2=104,196$ in our
application. So it is advisable to trim non-significant $\hat{\rho}%
_{jj^{\prime }}$s prior to clustering. A test for the hypothesis $%
H_{0,jj^{\prime }}:\rho _{jj^{\prime }}=0$ is the following (Seber 2004,
ch.~5.7.3): let $\{r_{k}^{2}\}$ be the $p=\min (p_{j},p_{j^{\prime }})$
non-zero eigenvalues of $\mathbf{S}_{jj}^{-1/2}\mathbf{S}_{jj^{\prime }}%
\mathbf{S}_{j^{\prime }j^{\prime }}^{-1}\mathbf{S}_{j^{\prime }j}\mathbf{S}%
_{jj}^{-1/2}$, or equivalently of $\mathbf{S}_{j^{\prime }j^{\prime }}^{-1/2}%
\mathbf{S}_{j^{\prime }j}\mathbf{S}_{jj}^{-1}\mathbf{S}_{jj^{\prime }}%
\mathbf{S}_{j^{\prime }j^{\prime }}^{-1/2}$, and $L=%
\prod_{k=1}^{p}(1-r_{k}^{2})$; then $Q_{jj^{\prime
}}=-\{n-1-(p_{j}+p_{j^{\prime }}+1)/2\}\log L$ is asymptotically $\chi _{\nu
}^{2}$ with $\nu =p_{j}p_{j^{\prime }}$ under the null hypothesis. To
determine non-significant $\hat{\rho}_{jj^{\prime }}$s at a simultaneous
level $\alpha $ we use Benjamini and Hochberg (1995) procedure: let $%
P_{jj^{\prime }}=P(\chi _{p^{2}}^{2}>Q_{jj^{\prime }})$ be the $p$-value for 
$H_{0,jj^{\prime }}$, and $\{P_{(k)}\}$ the set of these $p$-values sorted
in increasing order; then the correlations for which $P_{(k)}\leq \alpha
k/\{d(d-1)/2\}$ are considered significant. For the non-significant $\hat{%
\rho}_{jj^{\prime }}$s, we set $\hat{\rho}_{jj^{\prime }}=0$ and then
proceed to apply the linkage algorithm. Clusters, if there are any, can be
found from the dendrogram using standard techniques (see Izenman 2008,
ch.~12.3; Hastie \emph{et al.}, 2009, ch.~14.3.12). The consistency of the
clusters can be evaluated using measures such as the Davies--Bouldin index
(Davies and Bouldin, 1979) or the Dunn index (Dunn, 1974).

\section{Simulations\label{sec:Simulations}}

In this section we study the consistency of the estimators by simulation. We
simulated data from model (\ref{eq:log_lmb_model}) for $d=1$, since
estimation is done separately for each $j$. We considered three
distributional situations that will arise in the Divvy data analysis of
Section \ref{sec:Example}: component scores that \emph{(i)} are independent
and identically distributed, \emph{(ii)} follow a trend, and \emph{(iii)}
are autocorrelated. We also studied the effect of the expected number of
observations per replication, the baseline rate $\int_{a}^{b}\lambda
_{0}(t)dt$, which is determined by $\mu (t)$.

To this end we considered model (\ref{eq:log_lmb_model}) with $\mu (t)=\sin
(\pi t)+c$, $\phi _{1}(t)=\sqrt{2}\sin (\pi t)$ and $\phi _{2}(t)=\sqrt{2}%
\sin (2\pi t)$, for $t\in \lbrack 0,1]$. Since $\int_{0}^{1}\exp \{\sin (\pi
t)\}dt=1.98$, we took $c=\log 5$ and $c=\log 15$, which give approximate
baseline rates 10 and 30, respectively. The $u_{ik}$s were generated as
follows:

\begin{enumerate}
\item Independent: $u_{1k},\ldots ,u_{nk}$ were independent $N(0,\sigma
_{k}^{2})$ with $\sigma _{1}=.3\sqrt{.6}$ and $\sigma _{2}=.3\sqrt{.4}$,
respectively, so that the overall variance was $.09$, with the first
component accounting for 60\% of the variability. (The $u_{i1}$s were
independent of the $u_{i2}$s in all three scenarios, since the component
scores are uncorrelated across $k$s by definition).

\item With quadratic trend: let $s_{i}=-(i-n/2)^{2}$, for $i=1,\ldots ,n$.
Then $u_{i1}=\{(s_{i}-\bar{s})/\mathrm{sd}(s_{i})\}\sqrt{.75}\sigma
_{1}+z_{i}\sqrt{.25}\sigma _{1}$, with $z_{i}$s independent and identically
distributed $N(0,1)$, and $u_{i2}$s independent and identically distributed $%
N(0,\sigma _{2}^{2})$, with $\sigma _{1}$ and $\sigma _{2}$ as in Scenario
1. The variance of $u_{i1}$ is still $\sigma _{1}^{2}$, but 75\% of it now
comes from the quadratic trend.

\item Autocorrelated: the $u_{i1}$s followed the autoregressive model $%
u_{i1}=z_{i}\sigma _{e}$ for $i=1$ and $u_{i1}=\rho u_{i-1,1}+z_{i}\sigma
_{e}$ for $i=2,\ldots ,n$, with $z_{i}$s independent and identically
distributed $N(0,1)$, $\rho =.8$ and $\sigma _{e}=\sigma _{1}\sqrt{1-\rho
^{2}}$, so the variance of the $u_{i1}$s was $\sigma _{1}^{2}$ as in the
previous scenarios. The $u_{i2}$s were independent and identically
distributed $N(0,\sigma _{2}^{2})$, and we took $\sigma _{1}$ and $\sigma
_{2}$ as in Scenarios 1 and 2.
\end{enumerate}

To get an idea of the $m_{i}$s produced by these models, we generated a
sample of size 100 for each baseline rate, and observed $m_{i}$s between 4
and 20 for baseline rate 10 and between 22 and 44 for baseline rate 30. Four
sample sizes were considered for each scenario: $n=50$, $100$, $200$ and $%
400 $.

For estimation of the functional parameters we used a cubic B-spline basis
with five equally spaced knots in $(0,1)$, which has dimension $q=9$, large
enough for the smooth functions we are estimating. We chose subjective but
visually reasonable smoothing parameters $\xi _{1}=\xi _{2}=10^{-5}$.

%TCIMACRO{\TeXButton{B}{\begin{table}[tbp] \centering}}%
%BeginExpansion
\begin{table}[tbp] \centering%
%EndExpansion
\begin{tabular}{ccccccccc}
&  & \multicolumn{3}{c}{rate 10} &  & \multicolumn{3}{c}{rate 30} \\ 
$n$ & Param & bias & std & rmse &  & bias & std & rmse \\ 
&  &  &  &  &  &  &  &  \\ 
50 & $\mu $ & \multicolumn{1}{r}{.45} & \multicolumn{1}{r}{1.33} & 
\multicolumn{1}{r}{1.40} &  & \multicolumn{1}{r}{.50} & \multicolumn{1}{r}{
.88} & \multicolumn{1}{r}{1.01} \\ 
& $\phi _{1}$ & \multicolumn{1}{r}{.39} & \multicolumn{1}{r}{.71} & 
\multicolumn{1}{r}{.81} &  & \multicolumn{1}{r}{.29} & \multicolumn{1}{r}{.62
} & \multicolumn{1}{r}{.69} \\ 
& $\phi _{2}$ & \multicolumn{1}{r}{.61} & \multicolumn{1}{r}{.83} & 
\multicolumn{1}{r}{1.04} &  & \multicolumn{1}{r}{.36} & \multicolumn{1}{r}{
.69} & \multicolumn{1}{r}{.78} \\ 
&  &  &  &  &  &  &  &  \\ 
100 & $\mu $ & \multicolumn{1}{r}{.53} & \multicolumn{1}{r}{.95} & 
\multicolumn{1}{r}{1.08} &  & \multicolumn{1}{r}{.55} & \multicolumn{1}{r}{
.60} & \multicolumn{1}{r}{.82} \\ 
& $\phi _{1}$ & \multicolumn{1}{r}{.30} & \multicolumn{1}{r}{.59} & 
\multicolumn{1}{r}{.66} &  & \multicolumn{1}{r}{.14} & \multicolumn{1}{r}{.44
} & \multicolumn{1}{r}{.47} \\ 
& $\phi _{2}$ & \multicolumn{1}{r}{.41} & \multicolumn{1}{r}{.72} & 
\multicolumn{1}{r}{.83} &  & \multicolumn{1}{r}{.22} & \multicolumn{1}{r}{.51
} & \multicolumn{1}{r}{.56} \\ 
&  &  &  &  &  &  &  &  \\ 
200 & $\mu $ & \multicolumn{1}{r}{.52} & \multicolumn{1}{r}{.64} & 
\multicolumn{1}{r}{.82} &  & \multicolumn{1}{r}{.53} & \multicolumn{1}{r}{.43
} & \multicolumn{1}{r}{.68} \\ 
& $\phi _{1}$ & \multicolumn{1}{r}{.20} & \multicolumn{1}{r}{.48} & 
\multicolumn{1}{r}{.52} &  & \multicolumn{1}{r}{.10} & \multicolumn{1}{r}{.30
} & \multicolumn{1}{r}{.32} \\ 
& $\phi _{2}$ & \multicolumn{1}{r}{.35} & \multicolumn{1}{r}{.62} & 
\multicolumn{1}{r}{.71} &  & \multicolumn{1}{r}{.17} & \multicolumn{1}{r}{.36
} & \multicolumn{1}{r}{.40} \\ 
&  &  &  &  &  &  &  &  \\ 
400 & $\mu $ & \multicolumn{1}{r}{.53} & \multicolumn{1}{r}{.45} & 
\multicolumn{1}{r}{.69} &  & \multicolumn{1}{r}{.53} & \multicolumn{1}{r}{.30
} & \multicolumn{1}{r}{.61} \\ 
& $\phi _{1}$ & \multicolumn{1}{r}{.21} & \multicolumn{1}{r}{.34} & 
\multicolumn{1}{r}{.40} &  & \multicolumn{1}{r}{.08} & \multicolumn{1}{r}{.21
} & \multicolumn{1}{r}{.22} \\ 
& $\phi _{2}$ & \multicolumn{1}{r}{.34} & \multicolumn{1}{r}{.57} & 
\multicolumn{1}{r}{.66} & \multicolumn{1}{r}{} & \multicolumn{1}{r}{.16} & 
\multicolumn{1}{r}{.24} & \multicolumn{1}{r}{.29}%
\end{tabular}

\caption{Simulation Results. Bias, standard deviation and root mean squared errors of parameter estimators
for independent component scores (Scenario 1). Quantities for $\mu$ were multiplied by 10.}%
\label{tab:Table_1_iid}%
%TCIMACRO{\TeXButton{E}{\end{table}}}%
%BeginExpansion
\end{table}%
%EndExpansion

Tables \ref{tab:Table_1_iid} to \ref{tab:Table_3_autoreg} report the
results. For $\mu $ we defined $\mathrm{bias}=\Vert E(\hat{\mu})-\mu \Vert $%
, $\mathrm{std}=[E\{\Vert \hat{\mu}-E(\hat{\mu})\Vert ^{2}\}]^{1/2}$ and $%
\mathrm{rmse}=\{E(\Vert \hat{\mu}-\mu \Vert ^{2})\}^{1/2}$, where $\Vert
\cdot \Vert $ is the usual $L^{2}[0,1]$ norm. For the $\phi _{k}$s we could
not use these quantities because of the sign indetermination (a priori, it
is not possible to tell if $\hat{\phi}_{k}$ is estimating $\phi _{k}$ or $%
-\phi _{k}$), so we considered the bivariate estimators $\hat{\phi}_{k}(s)%
\hat{\phi}_{k}(t)$ of $\phi _{k}(s)\phi _{k}(t)$ instead, which are
sign-invariant, and defined bias, standard deviation and root mean squared
error as before, except that $\Vert \cdot \Vert $ was the bivariate $L^{2}$
norm on $[0,1]\times \lbrack 0,1]$. The expectations were approximated by
Monte Carlo based on 200 replications of each scenario.

Table \ref{tab:Table_1_iid} shows that, for independent and identically
distributed component scores, the estimators behave as expected: estimation
errors decrease as $n$ increases for each baseline rate, and they are lower
for the higher baseline rate. The bias of $\hat{\mu}$ does not decrease with 
$n$, but this is due to the suboptimal choice of smoothing parameter.

Table \ref{tab:Table_2_trend} shows the results for Scenario 2, where the
first component score follows a quadratic trend, and we see that they are
almost identical to those in Table \ref{tab:Table_1_iid}, so the estimators
work equally well in both situations. Table \ref{tab:Table_3_autoreg} shows
the results for Scenario 3, the autoregressive first component scores. The
mean squared errors of $\hat{\phi}_{1}$ and $\hat{\phi}_{2}$ are somewhat
larger than in the previous scenarios, but only by 20\% at most, and they
still decrease as $n$ increases, so the estimators are also consistent in
this scenario.

%TCIMACRO{\TeXButton{B}{\begin{table}[tbp] \centering}}%
%BeginExpansion
\begin{table}[tbp] \centering%
%EndExpansion
\begin{tabular}{ccccccccc}
&  & \multicolumn{3}{c}{rate 10} &  & \multicolumn{3}{c}{rate 30} \\ 
$n$ & Param & bias & std & rmse &  & bias & std & rmse \\ 
&  &  &  &  &  &  &  &  \\ 
50 & $\mu $ & \multicolumn{1}{r}{.46} & \multicolumn{1}{r}{1.29} & 
\multicolumn{1}{r}{1.37} &  & \multicolumn{1}{r}{.50} & \multicolumn{1}{r}{
.82} & \multicolumn{1}{r}{.96} \\ 
& $\phi _{1}$ & \multicolumn{1}{r}{.38} & \multicolumn{1}{r}{.70} & 
\multicolumn{1}{r}{.80} &  & \multicolumn{1}{r}{.25} & \multicolumn{1}{r}{.59
} & \multicolumn{1}{r}{.64} \\ 
& $\phi _{2}$ & \multicolumn{1}{r}{.60} & \multicolumn{1}{r}{.83} & 
\multicolumn{1}{r}{1.02} &  & \multicolumn{1}{r}{.35} & \multicolumn{1}{r}{
.69} & \multicolumn{1}{r}{.78} \\ 
&  & \multicolumn{7}{c}{} \\ 
100 & $\mu $ & \multicolumn{1}{r}{.48} & \multicolumn{1}{r}{.91} & 
\multicolumn{1}{r}{1.02} &  & \multicolumn{1}{r}{.50} & \multicolumn{1}{r}{
.59} & \multicolumn{1}{r}{.77} \\ 
& $\phi _{1}$ & \multicolumn{1}{r}{.28} & \multicolumn{1}{r}{.60} & 
\multicolumn{1}{r}{.66} &  & \multicolumn{1}{r}{.17} & \multicolumn{1}{r}{.48
} & \multicolumn{1}{r}{.51} \\ 
& $\phi _{2}$ & \multicolumn{1}{r}{.45} & \multicolumn{1}{r}{.75} & 
\multicolumn{1}{r}{.87} &  & \multicolumn{1}{r}{.24} & \multicolumn{1}{r}{.54
} & \multicolumn{1}{r}{.59} \\ 
&  & \multicolumn{7}{c}{} \\ 
200 & $\mu $ & \multicolumn{1}{r}{.50} & \multicolumn{1}{r}{.64} & 
\multicolumn{1}{r}{.82} &  & \multicolumn{1}{r}{.50} & \multicolumn{1}{r}{.40
} & \multicolumn{1}{r}{.64} \\ 
& $\phi _{1}$ & \multicolumn{1}{r}{.23} & \multicolumn{1}{r}{.46} & 
\multicolumn{1}{r}{.51} &  & \multicolumn{1}{r}{.10} & \multicolumn{1}{r}{.33
} & \multicolumn{1}{r}{.35} \\ 
& $\phi _{2}$ & \multicolumn{1}{r}{.34} & \multicolumn{1}{r}{.62} & 
\multicolumn{1}{r}{.71} &  & \multicolumn{1}{r}{.18} & \multicolumn{1}{r}{.38
} & \multicolumn{1}{r}{.42} \\ 
&  & \multicolumn{7}{c}{} \\ 
400 & $\mu $ & \multicolumn{1}{r}{.49} & \multicolumn{1}{r}{.45} & 
\multicolumn{1}{r}{.66} &  & \multicolumn{1}{r}{.52} & \multicolumn{1}{r}{.30
} & \multicolumn{1}{r}{.60} \\ 
& $\phi _{1}$ & \multicolumn{1}{r}{.18} & \multicolumn{1}{r}{.34} & 
\multicolumn{1}{r}{.39} &  & \multicolumn{1}{r}{.08} & \multicolumn{1}{r}{.22
} & \multicolumn{1}{r}{.23} \\ 
& $\phi _{2}$ & \multicolumn{1}{r}{.34} & \multicolumn{1}{r}{.55} & 
\multicolumn{1}{r}{.65} & \multicolumn{1}{r}{} & \multicolumn{1}{r}{.18} & 
\multicolumn{1}{r}{.31} & \multicolumn{1}{r}{.36}%
\end{tabular}

\caption{Simulation Results. Bias, standard deviation and root mean squared errors of parameter estimators
for component scores with a trend (Scenario 2). Quantities for $\mu$ were multiplied by 10.}%
\label{tab:Table_2_trend}%
%TCIMACRO{\TeXButton{E}{\end{table}}}%
%BeginExpansion
\end{table}%
%EndExpansion

%TCIMACRO{\TeXButton{B}{\begin{table}[tbp] \centering}}%
%BeginExpansion
\begin{table}[tbp] \centering%
%EndExpansion
\begin{tabular}{ccccccccc}
&  & \multicolumn{3}{c}{rate 10} &  & \multicolumn{3}{c}{rate 30} \\ 
$n$ & Param & bias & std & rmse &  & bias & std & rmse \\ 
&  & \multicolumn{7}{c}{} \\ 
50 & $\mu $ & \multicolumn{1}{r}{.46} & \multicolumn{1}{r}{1.56} & 
\multicolumn{1}{r}{1.63} &  & \multicolumn{1}{r}{.46} & \multicolumn{1}{r}{
1.21} & \multicolumn{1}{r}{1.29} \\ 
& $\phi _{1}$ & \multicolumn{1}{r}{.45} & \multicolumn{1}{r}{.75} & 
\multicolumn{1}{r}{.88} &  & \multicolumn{1}{r}{.45} & \multicolumn{1}{r}{.71
} & \multicolumn{1}{r}{.84} \\ 
& $\phi _{2}$ & \multicolumn{1}{r}{.60} & \multicolumn{1}{r}{.83} & 
\multicolumn{1}{r}{1.03} &  & \multicolumn{1}{r}{.50} & \multicolumn{1}{r}{
.77} & \multicolumn{1}{r}{.92} \\ 
&  & \multicolumn{7}{c}{} \\ 
100 & $\mu $ & \multicolumn{1}{r}{.45} & \multicolumn{1}{r}{1.12} & 
\multicolumn{1}{r}{1.21} &  & \multicolumn{1}{r}{.43} & \multicolumn{1}{r}{
.90} & \multicolumn{1}{r}{.99} \\ 
& $\phi _{1}$ & \multicolumn{1}{r}{.29} & \multicolumn{1}{r}{.63} & 
\multicolumn{1}{r}{.69} &  & \multicolumn{1}{r}{.21} & \multicolumn{1}{r}{.52
} & \multicolumn{1}{r}{.56} \\ 
& $\phi _{2}$ & \multicolumn{1}{r}{.48} & \multicolumn{1}{r}{.77} & 
\multicolumn{1}{r}{.91} &  & \multicolumn{1}{r}{.25} & \multicolumn{1}{r}{.57
} & \multicolumn{1}{r}{.62} \\ 
&  & \multicolumn{7}{c}{} \\ 
200 & $\mu $ & \multicolumn{1}{r}{.56} & \multicolumn{1}{r}{.83} & 
\multicolumn{1}{r}{1.00} &  & \multicolumn{1}{r}{.49} & \multicolumn{1}{r}{
.64} & \multicolumn{1}{r}{.81} \\ 
& $\phi _{1}$ & \multicolumn{1}{r}{.21} & \multicolumn{1}{r}{.50} & 
\multicolumn{1}{r}{.54} &  & \multicolumn{1}{r}{.12} & \multicolumn{1}{r}{.36
} & \multicolumn{1}{r}{.38} \\ 
& $\phi _{2}$ & \multicolumn{1}{r}{.36} & \multicolumn{1}{r}{.65} & 
\multicolumn{1}{r}{.74} &  & \multicolumn{1}{r}{.17} & \multicolumn{1}{r}{.40
} & \multicolumn{1}{r}{.44} \\ 
&  & \multicolumn{7}{c}{} \\ 
400 & $\mu $ & \multicolumn{1}{r}{.47} & \multicolumn{1}{r}{.56} & 
\multicolumn{1}{r}{.74} &  & \multicolumn{1}{r}{.52} & \multicolumn{1}{r}{.45
} & \multicolumn{1}{r}{.69} \\ 
& $\phi _{1}$ & \multicolumn{1}{r}{.20} & \multicolumn{1}{r}{.34} & 
\multicolumn{1}{r}{.39} &  & \multicolumn{1}{r}{.09} & \multicolumn{1}{r}{.25
} & \multicolumn{1}{r}{.26} \\ 
& $\phi _{2}$ & \multicolumn{1}{r}{.35} & \multicolumn{1}{r}{.54} & 
\multicolumn{1}{r}{.64} & \multicolumn{1}{r}{} & \multicolumn{1}{r}{.18} & 
\multicolumn{1}{r}{.30} & \multicolumn{1}{r}{.35}%
\end{tabular}

\caption{Simulation Results. Bias, standard deviation and root mean squared errors of parameter estimators
for autoregressive component scores (Scenario 3). Quantities for $\mu$ were multiplied by 10.}%
\label{tab:Table_3_autoreg}%
%TCIMACRO{\TeXButton{E}{\end{table}}}%
%BeginExpansion
\end{table}%
%EndExpansion

In addition to consistency of the parameter estimators, it is also important
to study the consistency of the component score estimators $\hat{u}_{ik}$s,
since they are used for inference (like clustering, in this paper). The
distance between the $\hat{u}_{ik}$s and the true $u_{ik}$s cannot be
measured directly, due to the sign indeterminacy, so we use the estimation
error of the variations $v_{ik}(t)=u_{ik}\phi _{k}(t)$ instead, which are
sign-invariant. We define the expected average error $\mathrm{eae}%
=E(\sum_{i=1}^{n}\Vert \hat{v}_{ik}-v_{ik}\Vert /n)$, where $\Vert \cdot
\Vert $ is the $L^{2}[0,1]$ norm. We also measure the association between
the $\hat{u}_{ik}$s and the $u_{ik}$s by the expected absolute correlation, $%
\mathrm{eac}=E\{|\mathrm{corr}(\hat{u}_{ik},u_{ik})|\}$, which is also
sign-invariant.

Table \ref{tab:Table_pred_err} shows the results. Again we see consistency,
an improvement in estimation as $n$ and/or the baseline rate increase, but
the latter has a bigger impact on the performance of the $\hat{u}_{ik}$s.
This was expected, since the $\hat{u}_{ik}$s can only use the $m_{i}$
observations available for replication $i$, whereas $\hat{\mu}(t)$ and the $%
\hat{\phi}_{k}(t)$s pool data across replications. Regarding the three
distributional scenarios, we see that there is almost no difference between
the independent identically distributed case and the model with quadratic
trend; the autoregressive model does show somewhat higher errors and lower
correlations than the other two, especially for $n=50$, but the difference
tends to vanish as $n$ increases. So we can say that the component score
estimators are consistent under the three scenarios.

When the component scores reveal a trend or autocorrelation, model (\ref%
{eq:log_lmb_model}) can be modified to accommodate such relationships, and
re-estimated. Other covariates on which the $u_{ikj}$s may depend can also
be incorporated. However, a detailed elaboration of these possibilities goes
beyond the scope of this paper.

%TCIMACRO{\TeXButton{B}{\begin{table}[tbp] \centering}}%
%BeginExpansion
\begin{table}[tbp] \centering%
%EndExpansion
\begin{tabular}{ccccccrccccrcccc}
&  & \multicolumn{4}{c}{Independent} &  & \multicolumn{4}{c}{With trend} & 
& \multicolumn{4}{c}{Autoregressive} \\ 
&  & \multicolumn{2}{c}{rate 10} & \multicolumn{2}{c}{rate 30} &  & 
\multicolumn{2}{c}{rate 10} & \multicolumn{2}{c}{rate 30} & 
\multicolumn{1}{c}{} & \multicolumn{2}{c}{rate 10} & \multicolumn{2}{c}{rate
30} \\ 
$n$ & Score & eae & eac & eae & eac & \multicolumn{1}{c}{} & eae & eac & eae
& eac & \multicolumn{1}{c}{} & eae & eac & eae & eac \\ 
&  & \multicolumn{1}{r}{} & \multicolumn{1}{r}{} & \multicolumn{1}{r}{} & 
\multicolumn{1}{r}{} &  & \multicolumn{1}{r}{} & \multicolumn{1}{r}{} & 
\multicolumn{1}{r}{} & \multicolumn{1}{r}{} &  & \multicolumn{1}{r}{} & 
\multicolumn{1}{r}{} & \multicolumn{1}{r}{} & \multicolumn{1}{r}{} \\ 
50 & pc 1 & \multicolumn{1}{r}{.28} & \multicolumn{1}{r}{.51} & 
\multicolumn{1}{r}{.18} & \multicolumn{1}{r}{.73} &  & \multicolumn{1}{r}{.28
} & \multicolumn{1}{r}{.52} & \multicolumn{1}{r}{.17} & \multicolumn{1}{r}{
.75} &  & \multicolumn{1}{r}{.29} & \multicolumn{1}{r}{.47} & 
\multicolumn{1}{r}{.20} & \multicolumn{1}{r}{.63} \\ 
& pc 2 & \multicolumn{1}{r}{.26} & \multicolumn{1}{r}{.33} & 
\multicolumn{1}{r}{.17} & \multicolumn{1}{r}{.59} &  & \multicolumn{1}{r}{.26
} & \multicolumn{1}{r}{.34} & \multicolumn{1}{r}{.17} & \multicolumn{1}{r}{
.60} &  & \multicolumn{1}{r}{.26} & \multicolumn{1}{r}{.34} & 
\multicolumn{1}{r}{.19} & \multicolumn{1}{r}{.51} \\ 
&  & \multicolumn{1}{r}{} & \multicolumn{1}{r}{} & \multicolumn{1}{r}{} & 
\multicolumn{1}{r}{} &  & \multicolumn{1}{r}{} & \multicolumn{1}{r}{} & 
\multicolumn{1}{r}{} & \multicolumn{1}{r}{} &  & \multicolumn{1}{r}{} & 
\multicolumn{1}{r}{} & \multicolumn{1}{r}{} & \multicolumn{1}{r}{} \\ 
100 & pc 1 & \multicolumn{1}{r}{.26} & \multicolumn{1}{r}{.54} & 
\multicolumn{1}{r}{.15} & \multicolumn{1}{r}{.77} &  & \multicolumn{1}{r}{.26
} & \multicolumn{1}{r}{.54} & \multicolumn{1}{r}{.16} & \multicolumn{1}{r}{
.77} &  & \multicolumn{1}{r}{.26} & \multicolumn{1}{r}{.52} & 
\multicolumn{1}{r}{.17} & \multicolumn{1}{r}{.73} \\ 
& pc 2 & \multicolumn{1}{r}{.24} & \multicolumn{1}{r}{.40} & 
\multicolumn{1}{r}{.15} & \multicolumn{1}{r}{.67} &  & \multicolumn{1}{r}{.24
} & \multicolumn{1}{r}{.39} & \multicolumn{1}{r}{.15} & \multicolumn{1}{r}{
.65} &  & \multicolumn{1}{r}{.25} & \multicolumn{1}{r}{.37} & 
\multicolumn{1}{r}{.16} & \multicolumn{1}{r}{.63} \\ 
&  & \multicolumn{1}{r}{} & \multicolumn{1}{r}{} & \multicolumn{1}{r}{} & 
\multicolumn{1}{r}{} &  & \multicolumn{1}{r}{} & \multicolumn{1}{r}{} & 
\multicolumn{1}{r}{} & \multicolumn{1}{r}{} &  & \multicolumn{1}{r}{} & 
\multicolumn{1}{r}{} & \multicolumn{1}{r}{} & \multicolumn{1}{r}{} \\ 
200 & pc 1 & \multicolumn{1}{r}{.24} & \multicolumn{1}{r}{.56} & 
\multicolumn{1}{r}{.14} & \multicolumn{1}{r}{.79} &  & \multicolumn{1}{r}{.24
} & \multicolumn{1}{r}{.57} & \multicolumn{1}{r}{.14} & \multicolumn{1}{r}{
.79} &  & \multicolumn{1}{r}{.24} & \multicolumn{1}{r}{.54} & 
\multicolumn{1}{r}{.15} & \multicolumn{1}{r}{.77} \\ 
& pc 2 & \multicolumn{1}{r}{.23} & \multicolumn{1}{r}{.43} & 
\multicolumn{1}{r}{.14} & \multicolumn{1}{r}{.69} &  & \multicolumn{1}{r}{.23
} & \multicolumn{1}{r}{.42} & \multicolumn{1}{r}{.14} & \multicolumn{1}{r}{
.68} &  & \multicolumn{1}{r}{.23} & \multicolumn{1}{r}{.42} & 
\multicolumn{1}{r}{.14} & \multicolumn{1}{r}{.68} \\ 
&  & \multicolumn{1}{r}{} & \multicolumn{1}{r}{} & \multicolumn{1}{r}{} & 
\multicolumn{1}{r}{} &  & \multicolumn{1}{r}{} & \multicolumn{1}{r}{} & 
\multicolumn{1}{r}{} & \multicolumn{1}{r}{} &  & \multicolumn{1}{r}{} & 
\multicolumn{1}{r}{} & \multicolumn{1}{r}{} & \multicolumn{1}{r}{} \\ 
400 & pc 1 & \multicolumn{1}{r}{.23} & \multicolumn{1}{r}{.58} & 
\multicolumn{1}{r}{.13} & \multicolumn{1}{r}{.80} &  & \multicolumn{1}{r}{.23
} & \multicolumn{1}{r}{.57} & \multicolumn{1}{r}{.13} & \multicolumn{1}{r}{
.80} &  & \multicolumn{1}{r}{.23} & \multicolumn{1}{r}{.57} & 
\multicolumn{1}{r}{.14} & \multicolumn{1}{r}{.79} \\ 
& pc 2 & \multicolumn{1}{r}{.22} & \multicolumn{1}{r}{.43} & 
\multicolumn{1}{r}{.13} & \multicolumn{1}{r}{.70} &  & \multicolumn{1}{r}{.22
} & \multicolumn{1}{r}{.43} & \multicolumn{1}{r}{.13} & \multicolumn{1}{r}{
.68} &  & \multicolumn{1}{r}{.22} & \multicolumn{1}{r}{.43} & 
\multicolumn{1}{r}{.13} & \multicolumn{1}{r}{.69}%
\end{tabular}

\caption{Simulation Results. Expected average error (eae) and expected absolute correlation (eac)
of component score estimators under the three scenarios of 
Tables $\ref{tab:Table_1_iid}$--$\ref{tab:Table_3_autoreg}$.}\label%
{tab:Table_pred_err}%
%TCIMACRO{\TeXButton{E}{\end{table}}}%
%BeginExpansion
\end{table}%
%EndExpansion

\section{Application: Chicago's Divvy bike sharing system\label{sec:Example}}

As mentioned in the Introduction, we analyze in this section the checkout
times of bike trips that took place between April 1 and November 31 of 2016
in Chicago's Divvy system. First, we fitted model (\ref{eq:log_lmb_model})
for the 458 bike stations that were active during this period. As spline
basis for the functional parameters we used cubic B-splines with ten equally
spaced knots in $(0,24)$. We fitted models with $p=6$ components, which were
sufficient to capture the most important modes of variation in the data and
can be estimated without inconvenient for most stations; only for station
386, the station with the lowest annual count (29 for the whole year), the
model could not be fitted due to insufficient data.

It is clearly infeasible to visually inspect the results for all stations,
but as an illustration we will analyze in more detail the results for
station 166, the station with median annual count. The estimated baseline
intensity function $\hat{\lambda}_{0,166}$ is shown in Figure \ref%
{fig:lmb0_166}. We see that $\hat{\lambda}_{0,166}$ has three peaks: the
first and largest one occurs at 7:30am, the second and smallest one at 1pm,
and the third one at 5:30pm. The integral of $\hat{\lambda}_{0,166}$ over $%
[0,24]$ is $17.66$, very close to the mean daily count of $17.64$, as
expected.

\FRAME{ftbpFU}{3.3892in}{2.5512in}{0pt}{\Qcb{Baseline intensity function of
daily bike demand for Divvy station 166, located at the intersection of
Wrightwood and Ashland avenues.}}{\Qlb{fig:lmb0_166}}{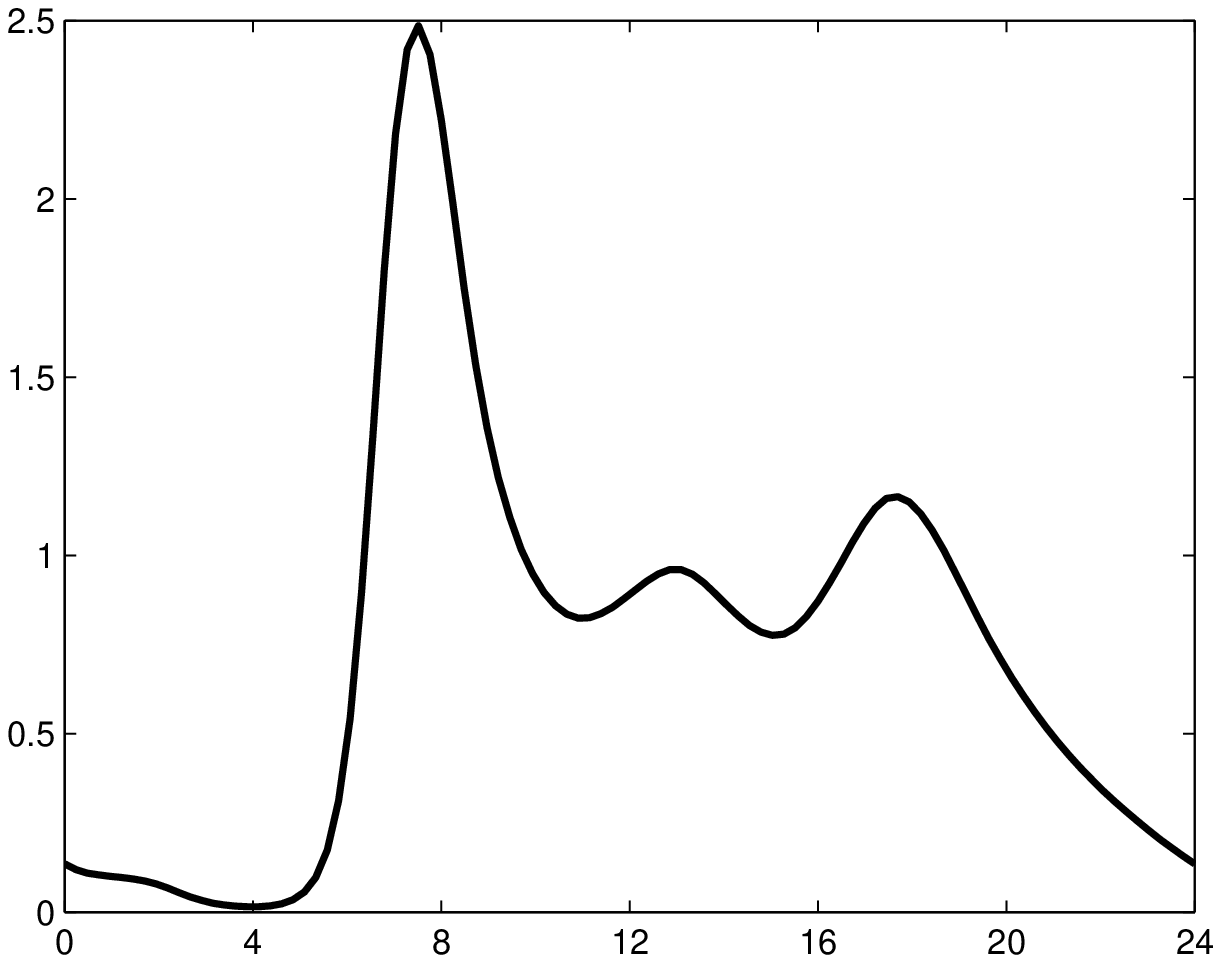}{\special%
{language "Scientific Word";type "GRAPHIC";maintain-aspect-ratio
TRUE;display "ICON";valid_file "F";width 3.3892in;height 2.5512in;depth
0pt;original-width 5.8219in;original-height 4.3708in;cropleft "0";croptop
"1";cropright "1";cropbottom "0";filename 'lmb0_166.eps';file-properties
"XNPEU";}}

To interpret the components $\hat{\psi}_{k,j}$ it is instructive to plot the
baseline function $\hat{\lambda}_{0j}$ alongside $\hat{\lambda}_{0j}^{+}=%
\hat{\lambda}_{0j}\hat{\psi}_{kj}^{c}$ and $\hat{\lambda}_{0j}^{-}=\hat{%
\lambda}_{0j}\hat{\psi}_{kj}^{-c}$, for some positive constant $c$ chosen
for convenient visualization (here we take it as twice the standard
deviation of the corresponding $\hat{u}_{ikj}$s). For the first component,
this is shown in Figure \ref{fig:pc1_166}(a). In Figure \ref{fig:pc1_166}(b)
we plotted the corresponding component scores $\hat{u}_{i,1,166}$ as a time
series on the index $i$. Figure \ref{fig:pc1_166}(a) shows that a negative
score corresponds to a sharpening of the morning peak and a positive score
corresponds to a flattening of this peak. This corresponds to weekday versus
weekend patterns of demand, respectively, as corroborated by Figure \ref%
{fig:pc1_166}(b), which shows a steady weekly periodicity (the
autocorrelation at lag 7 is $.68$), with peaks occurring almost always on
Sundays and troughs mostly on Thursdays or Wednesdays. In Figure \ref%
{fig:week_166} we show the $244$ estimated daily intensity functions,
separating weekdays (Figure \ref{fig:week_166}(a)) from weekends (Figure \ref%
{fig:week_166}(b)); the absence of the morning peaks in Figure \ref%
{fig:week_166}(b) is clear.

\FRAME{ftbpFU}{5.3714in}{2.5512in}{0pt}{\Qcb{First multiplicative component
of daily bike demand for Divvy station 166. (a) Baseline (solid line) and
baseline multiplied by a positive (dotted line) and negative (dashed line)
exponent of the component. (b) Daily component scores as a time series.}}{%
\Qlb{fig:pc1_166}}{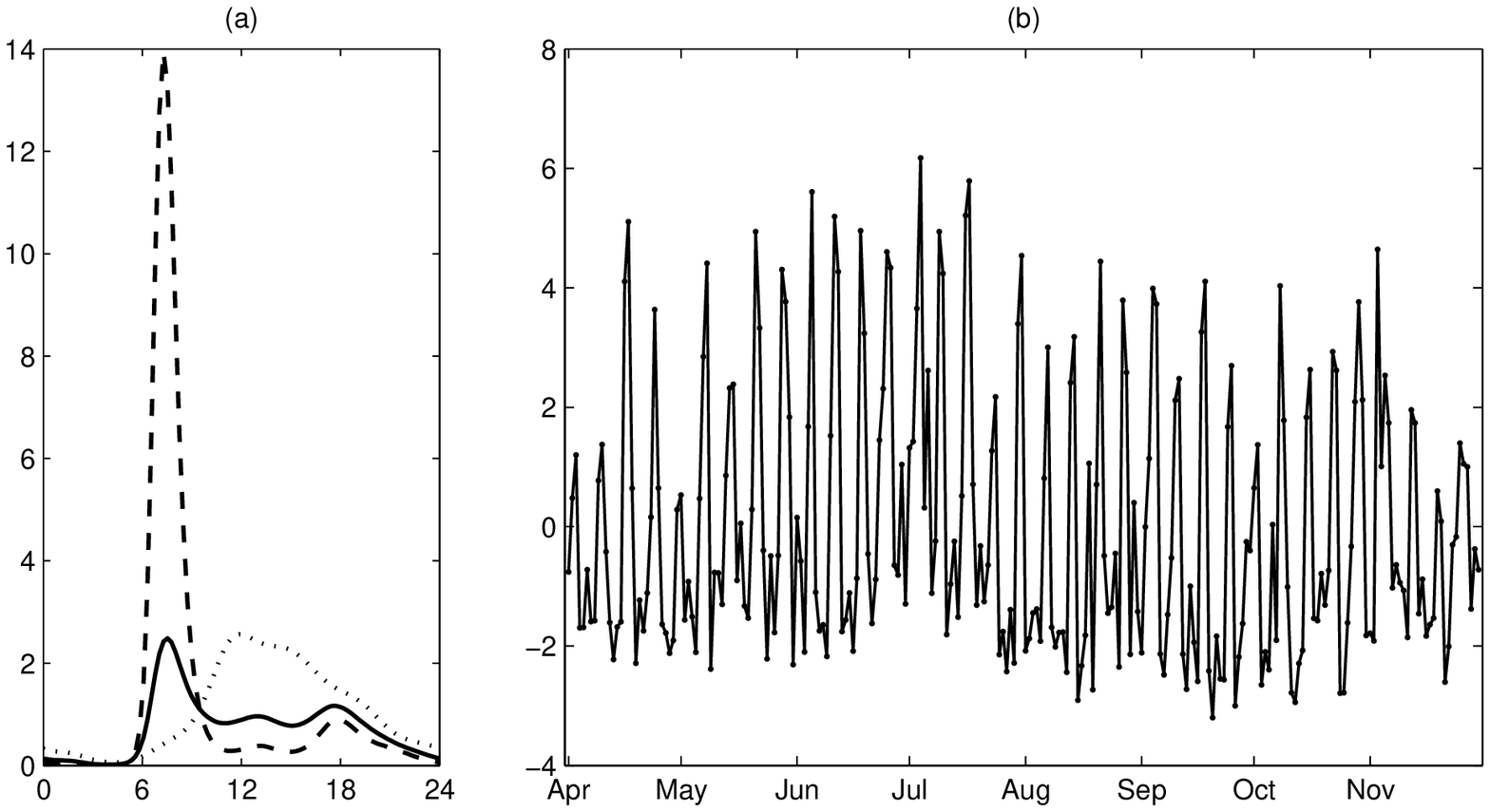}{\special{language "Scientific Word";type
"GRAPHIC";maintain-aspect-ratio TRUE;display "ICON";valid_file "F";width
5.3714in;height 2.5512in;depth 0pt;original-width 9.2353in;original-height
4.3613in;cropleft "0";croptop "1";cropright "1";cropbottom "0";filename
'pc1_166.eps';file-properties "XNPEU";}}

\FRAME{ftbpFU}{5.943in}{2.5512in}{0pt}{\Qcb{Daily intensity functions of
bike demand for Divvy station 166, (a) weekdays, (b) weekends.}}{\Qlb{%
fig:week_166}}{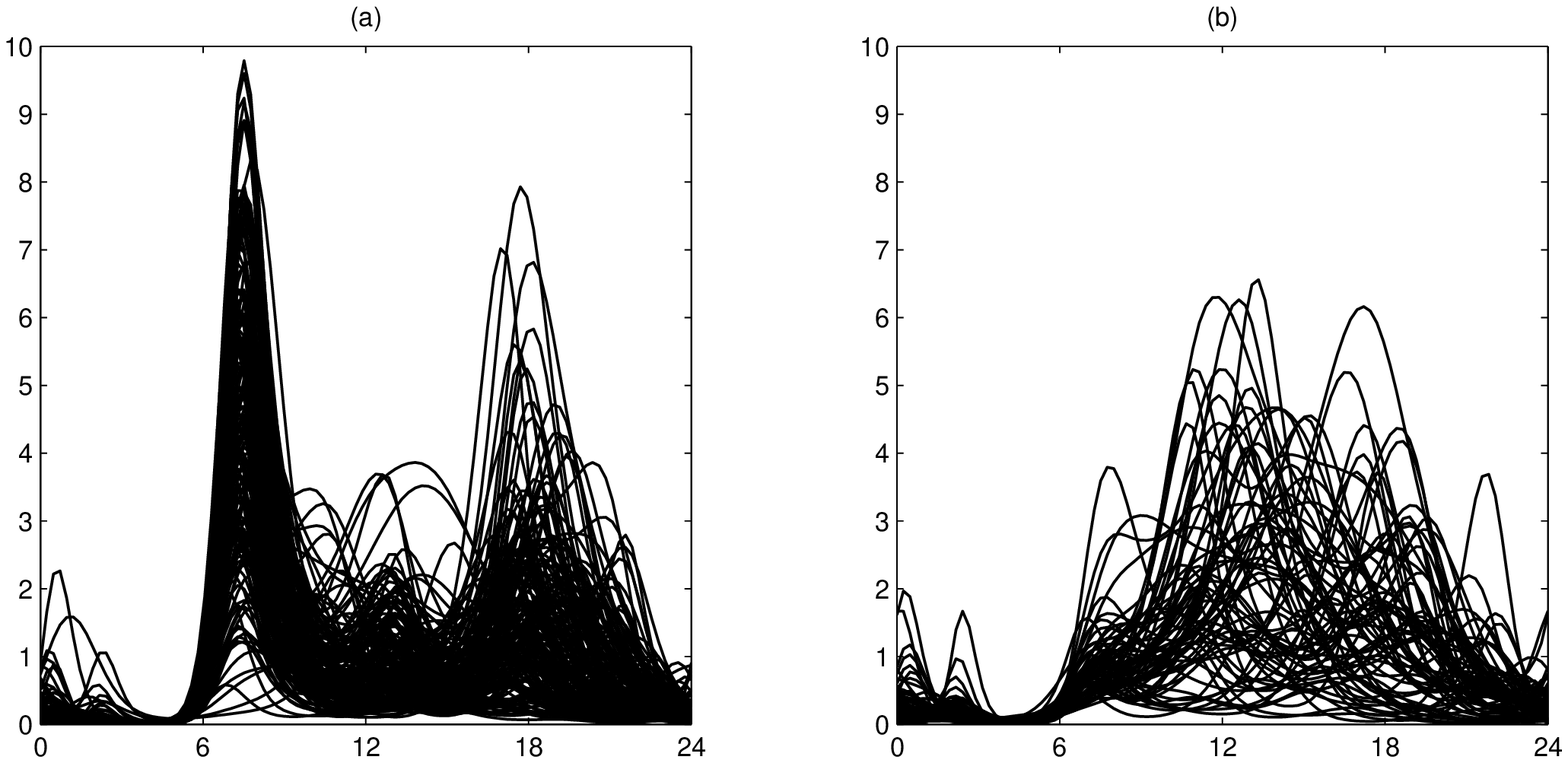}{\special{language "Scientific Word";type
"GRAPHIC";maintain-aspect-ratio TRUE;display "ICON";valid_file "F";width
5.943in;height 2.5512in;depth 0pt;original-width 10.2229in;original-height
4.3613in;cropleft "0";croptop "1";cropright "1";cropbottom "0";filename
'wkday_wkend_166.eps';file-properties "XNPEU";}}

The second component (Figure \ref{fig:pc2_166}(a)) explains overall count
variation. Overall bike usage is strongly seasonal, as shown in Figure \ref%
{fig:pc2_166}(b), with demand increasing from early Spring to Summer (the
maximum occurs in June) and decreasing thereafter. The rest of the
components explain finer-detailed aspects of bike demand.

\FRAME{ftbpFU}{5.3714in}{2.5512in}{0pt}{\Qcb{Second multiplicative component
of daily bike demand for Divvy station 166. (a) Baseline (solid line) and
baseline multiplied by a positive (dotted line) and negative (dashed line)
exponent of the component. (b) Daily component scores as a time series.}}{%
\Qlb{fig:pc2_166}}{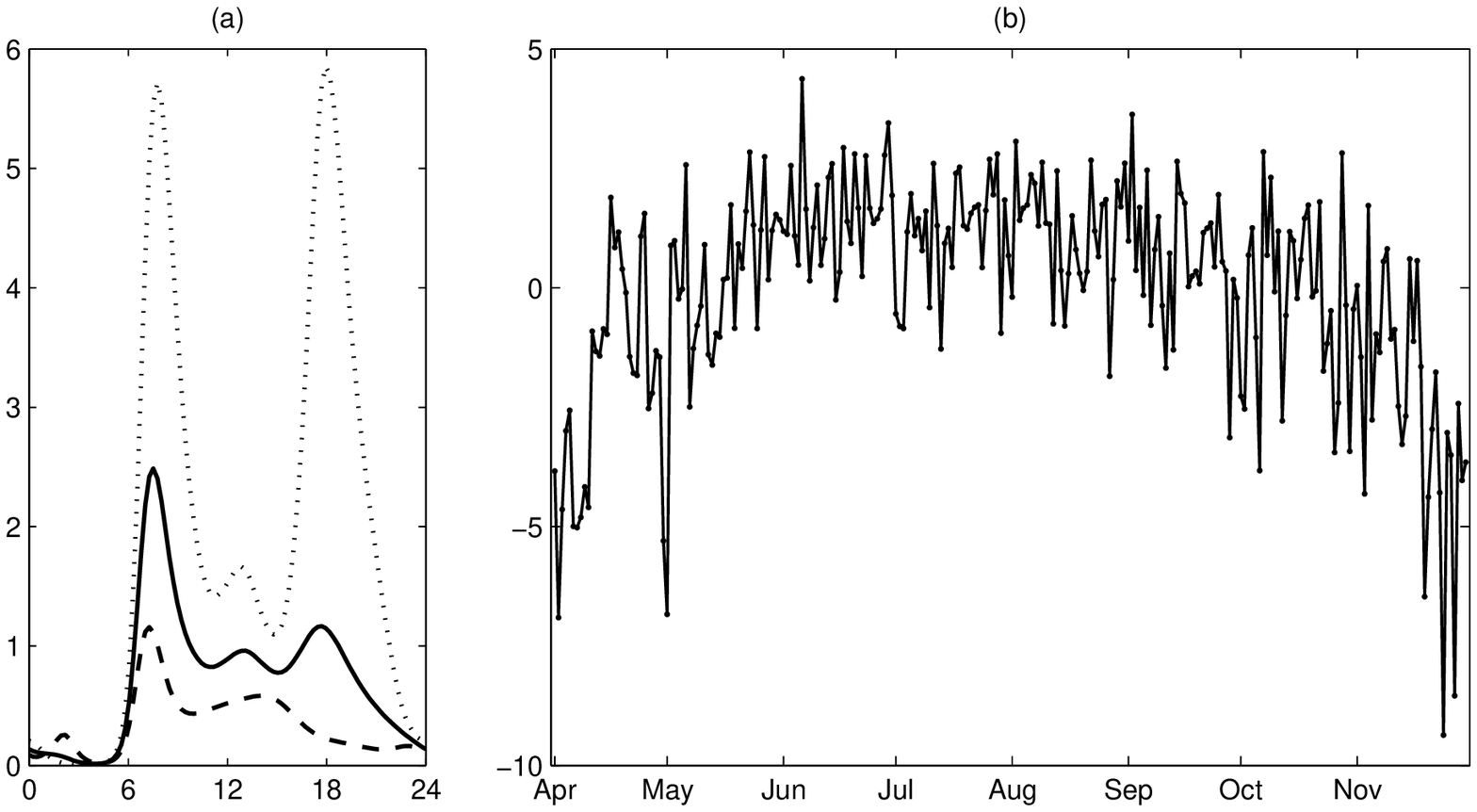}{\special{language "Scientific Word";type
"GRAPHIC";maintain-aspect-ratio TRUE;display "ICON";valid_file "F";width
5.3714in;height 2.5512in;depth 0pt;original-width 9.2353in;original-height
4.3613in;cropleft "0";croptop "1";cropright "1";cropbottom "0";filename
'pc2_166.eps';file-properties "XNPEU";}}

After fitting model (\ref{eq:lmb_model}) for all bike stations, we computed
the canonical correlations (\ref{eq:rho_hat}) for all pairs. The largest one
turned out to be $.98$ and the smallest one $.17$. The largest correlation
corresponds to bike stations 75 and 91, located at the main entrances of
Union and Ogilvy train stations, respectively. Although these bike stations
are relatively close to each other (556 m, four city blocks), they are not
the closest. For example, station 73 is closer to station 75 (277 m, two
city blocks) but their correlation is lower ($.90$), and station 169 is 452
m (three city blocks) away from station 75, closer than Ogilvy is but in the
opposite direction and without any train stations nearby, so their
correlation is only $.72$. It is clear, then, that correlations are not
functions of distance alone but also of type of usage; the spatial
correlations are not isotropic.

\FRAME{ftbpFU}{4.6847in}{3.5232in}{0pt}{\Qcb{Dendrogram of complete-linkage
clustering of bike stations in the Divvy system.}}{\Qlb{fig:dendro}}{%
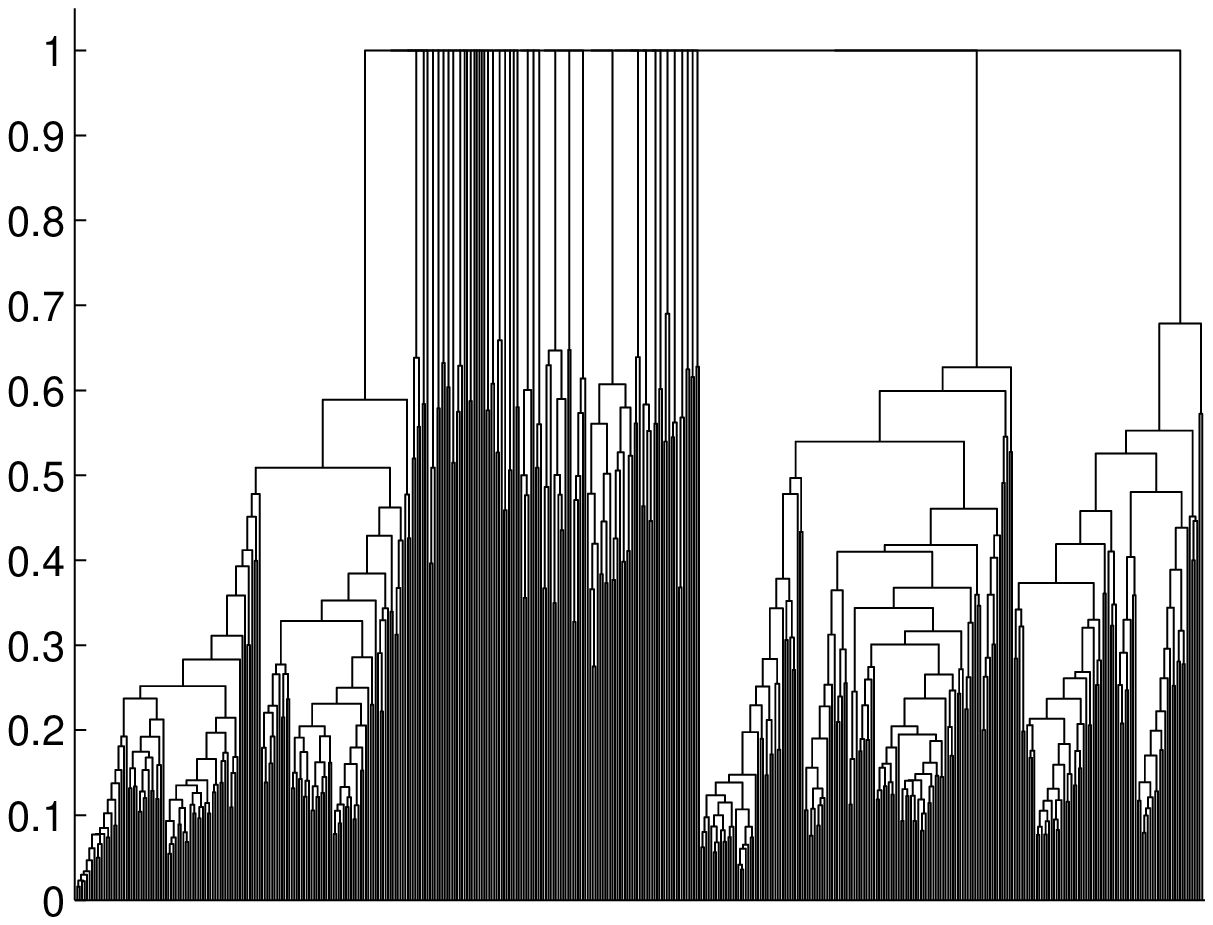}{\special{language "Scientific Word";type
"GRAPHIC";maintain-aspect-ratio TRUE;display "ICON";valid_file "F";width
4.6847in;height 3.5232in;depth 0pt;original-width 5.8219in;original-height
4.3708in;cropleft "0";croptop "1";cropright "1";cropbottom "0";filename
'dendrogram.eps';file-properties "XNPEU";}}

Then it is instructive to apply clustering methods to the correlations and
try to associate the clusters with different patterns of usage. The
clustering procedure of Section \ref{sec:Clustering} gives the dendrogram
shown in Figure \ref{fig:dendro}. The vertical axis of the dendrogram
indicates the distance of the objects being connected. Three big clusters
are discernible in Figure \ref{fig:dendro}, with a maximum distance of about 
$.70$, so the correlations of bike stations within the clusters are at least 
$.30$. These clusters include 136, 127 and 77 bike stations respectively, so
they account for 340 of the 458 bike stations in the system (most of the
others had non-significant correlations that were trimmed as explained in
Section \ref{sec:Clustering}).

\FRAME{ftbpFU}{6.4671in}{2.4725in}{0pt}{\Qcb{Clusters of bike stations in
the Divvy system. (a) Largest cluster, 136 stations; (b) second largest
cluster, 127 stations, (c) third largest cluster, 77 stations.}}{\Qlb{%
fig:cl_points}}{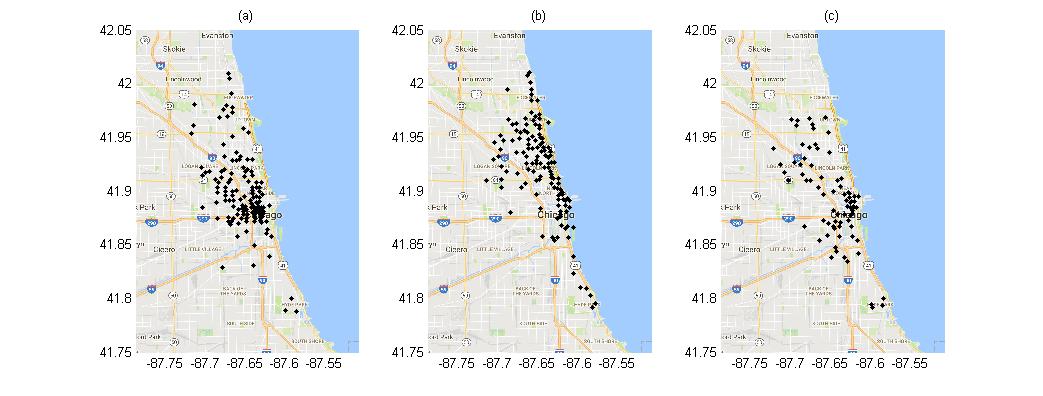}{\special{language "Scientific Word";type
"GRAPHIC";maintain-aspect-ratio TRUE;display "ICON";valid_file "F";width
6.4671in;height 2.4725in;depth 0pt;original-width 10.8646in;original-height
4.1252in;cropleft "0";croptop "1";cropright "1";cropbottom "0";filename
'cluster_points.jpg';file-properties "XNPEU";}}

\FRAME{ftbpFU}{5.8211in}{2.5503in}{0pt}{\Qcb{Baseline density functions for
the three clusters of bike stations in Figure \protect\ref{fig:cl_points}.}}{%
\Qlb{fig:cl_pdfs}}{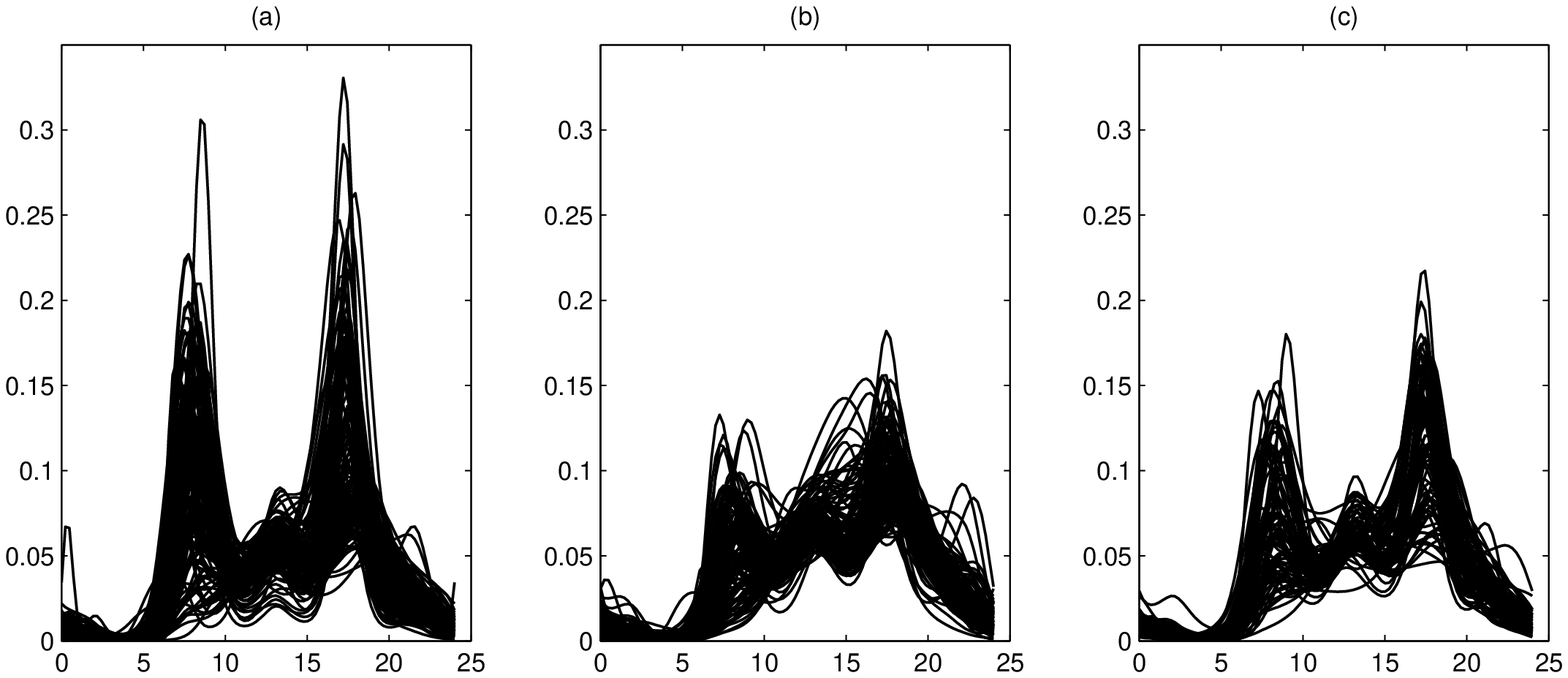}{\special{language "Scientific
Word";type "GRAPHIC";maintain-aspect-ratio TRUE;display "ICON";valid_file
"F";width 5.8211in;height 2.5503in;depth 0pt;original-width
8.5867in;original-height 3.7395in;cropleft "0";croptop "1";cropright
"1";cropbottom "0";filename 'cluster_densities.eps';file-properties "XNPEU";}%
}

The locations of stations in each cluster are shown in Figure \ref%
{fig:cl_points}. We also show the baseline density functions for each
station, $\tilde{\lambda}_{0j}=\lambda _{0j}/\int \lambda _{0j}$, in Figure %
\ref{fig:cl_pdfs}. Although we are clustering by correlation and not by
distance between baselines, the baseline densities do help interpret the
type of usage given to the stations in each cluster. We see in Figure \ref%
{fig:cl_pdfs}(a) that most densities in this cluster show a typical
weekday-usage pattern (compare with Figure \ref{fig:week_166}(a)); Figure %
\ref{fig:cl_points}(a) shows that most stations in downtown Chicago, and
specifically in \textquotedblleft the Loop\textquotedblright , belong to
this cluster, so Cluster 1 consists of bike stations that are mostly used
for commute. The densities in Figure \ref{fig:cl_pdfs}(b) show a
weekend-usage pattern (compare with Figure \ref{fig:week_166}(b)), and
Figure \ref{fig:cl_points}(b) shows that most stations along the lake shore
belong to this cluster, so Cluster 2 consists of stations that are mostly
used for leisure trips. The third cluster is somewhere in between.

\section*{Acknowledgement}

This research was partly supported by US National Science Foundation grant
DMS 1505780.

\section{Appendix: Computation of functional canonical correlations}

Let $U(t)$ and $V(t)$ be two stochastic processes admitting finite
expansions $U(t)=\mu _{U}(t)+\sum_{k=1}^{p}u_{k}\phi _{k}(t)$ and $V(t)=\mu
_{V}(t)+\sum_{k=1}^{q}v_{k}\psi _{k}(t)$, where the $\phi _{k}$s and the $%
\psi _{k}$s are orthonormal. The canonical correlation coefficient is $\rho
=\max_{\alpha ,\beta }\func{corr}(\langle \alpha ,U\rangle ,\langle \beta
,V\rangle )$, where $\alpha $ and $\beta $ are arbitrary square-integrable
functions. Any $\alpha $ and $\beta $ can be decomposed as $\alpha
(t)=\sum_{k=1}^{p}a_{k}\phi _{k}(t)+\eta (t)$, with $\eta $ orthogonal to
the $\phi _{k}$s, and $\beta (t)=\sum_{k=1}^{q}b_{k}\psi _{k}(t)+\xi (t)$,
with $\xi $ orthogonal to the $\psi _{k}$s. Then $\langle \alpha ,U-\mu
_{U}\rangle =\sum_{k=1}^{p}a_{k}u_{k}$ and $\langle \beta ,V-\mu _{V}\rangle
=\sum_{k=1}^{q}b_{k}v_{k}$. Let $\mathbf{a}=(a_{1},\ldots ,a_{p})$, $\mathbf{%
U}=(U_{1},\ldots ,U_{p})$, $\mathbf{b}=(b_{1},\ldots ,b_{q})$ and $\mathbf{V}%
=(V_{1},\ldots ,V_{q})$; then $\langle \alpha ,U-\mu _{U}\rangle =\mathbf{a}%
^{T}\mathbf{U}$ and $\langle \beta ,V-\mu _{V}\rangle =\mathbf{b}^{T}\mathbf{%
V}$. Since $\func{corr}(\langle \alpha ,U\rangle ,\langle \beta ,V\rangle )=%
\func{corr}(\langle \alpha ,U-\mu _{U}\rangle ,\langle \beta ,V-\mu
_{V}\rangle )$, then $\rho =\max_{\mathbf{a},\mathbf{b}}\func{corr}(\mathbf{a%
}^{T}\mathbf{U},\mathbf{b}^{T}\mathbf{V})$, which is the standard
multivariate canonical correlation coefficient for $\mathbf{U}$ and $\mathbf{%
V}$. In particular, if $U(t)=\log \Lambda _{j}(t)$ and $V(t)=\log \Lambda
_{j^{\prime }}(t)$, where $\Lambda _{j}(t)$ is the random function that
generates the $\lambda _{ij}(t)$s in model (\ref{eq:log_lmb_model}), we have
that $\mathbf{U}$ are the component scores for site $j$ and $\mathbf{V}$ are
the component scores for site $j^{\prime }$.

\section*{References}

\begin{description}
\item Baddeley, A. (2007). Spatial point processes and their applications.
In \emph{Stochastic Geometry}, Lecture Notes in Mathematics 1892, pp.~1--75.
Springer, New York.

\item Baddeley, B. A., Moyeed, R. A., Howard, C. V., and Boyde, A. (1993).
Analysis of three-dimensional point pattern with replication. \emph{Applied
Statistics} \textbf{42} 641--668.

\item Baddeley, A., Rubak, E., and Turner, R. (2015). \emph{Spatial Point
Patterns: Methodology and Applications with R}. CRC Press, Boca Raton, USA.

\item Benjamini, Y., and Hochberg, Y. (1995). Controlling the false
discovery rate: A practical and powerful approach to multiple testing. \emph{%
Journal of the Royal Statistical Society Series B} \textbf{57} 289--300.

\item Borgnat, P., Robardet, C., Rouquier, J., Abry, P., Flandrin, P., and
Fleury, E. (2011). Shared bicycles in a city: A signal processing and data
analysis perspective. \emph{Advances in Complex Systems} \textbf{14} 1--24.

\item Bouzas, P.R., and Ruiz-Fuentes, N. (2015). A review on functional data
analysis for Cox processes. \emph{Bolet\'{\i}n de Estad\'{\i}stica e
Investigaci\'{o}n Operativa }\textbf{31} 215--230.

\item Davies, D., and Bouldin, D. (1979). A cluster separation measure. 
\emph{IEEE Transactions on Pattern Analysis and Machine Intelligence} 
\textbf{1} 224--227.

\item De Boor, C. (1978). \emph{A Practical Guide to Splines}. Springer, New
York.

\item Delicado, P., Giraldo, R., Comas, C., and Mateu, J. (2010). Statistics
for spatial functional data: Some recent contributions. \emph{Environmetrics 
}\textbf{21} 224--239.

\item Diggle, P. (2013). \emph{Statistical Analysis of Spatial and
Spatio-Temporal Point Patterns, Third Edition}. CRC Press, Boca Raton, USA.

\item Diggle, P. J., Lange, N. and Bene\v{s}, F. M. (1991). Analysis of
variance for replicated spatial point patterns in clinical neuroanatomy. 
\emph{Journal of the American Statistical Association} \textbf{86} 618--625.

\item Diggle, P. J., Mateu, J. and Clough, H. E. (2000). A comparison
between parametric and non-parametric approaches to the analysis of
replicated spatial point pattern. \emph{Advances in Applied Probability} 
\textbf{32} 331--343.

\item Dunn, J. (1974). Well separated clusters and optimal fuzzy partitions. 
\emph{Journal of Cybernetics} \textbf{4} 95--104.

\item Eilers, P.H.C., and Marx, B.D. (1996). Flexible smoothing with
B-splines and penalties (with discussion). \emph{Statistical Science }%
\textbf{11} 89--121.

\item Gervini, D. (2016). Independent component models for replicated point
processes. \emph{Spatial Statistics} \textbf{18} 474-488.

\item Hastie, T., Tibshirani, R., and Friedman, J. (2009). \emph{The
Elements of Statistical Learning. Data Mining, Inference, and Prediction.
Second Edition. }Springer, New York.

\item Horv\'{a}th, L., and Kokoszka, P. (2012). \emph{Inference for
Functional Data with Applications.} Springer, New York.

\item Izenman, A.J. (2008). \emph{Modern Multivariate Statistical
Techniques. Regression, Classification and Manifold Learning.} Springer, New
York.

\item Jolliffe, I.T. (2002). \emph{Principal Component Analysis. Second
Edition.} Springer, New York.

\item Landau, S., Rabe-Hesketh, S., and Everall, I.P. (2004). Nonparametric
one-way analysis of variance of replicated bivariate spatial point patterns. 
\emph{Biometrical Journal} \textbf{46} 19--34.

\item Li, Y., and Guan, Y. (2014). Functional principal component analysis
of spatiotemporal point processes with applications in disease surveillance. 
\emph{Journal of the American Statistical Association }\textbf{109}
1205--1215.

\item Mateu, J. (2001). Parametric procedures in the analysis of replicated
pairwise interaction point patterns. \emph{Biometrical Journal} \textbf{43}
375--394.

\item Menafoglio, A., and Secchi, P. (2017). Statistical analysis of complex
and spatially dependent data: A review of object oriented spatial
statistics. \emph{European Journal of Operational Research} \textbf{258}
401--410.

\item M\o ller, J., and Waagepetersen, R.P. (2004). \emph{Statistical
Inference and Simulation for Spatial Point Processes}. Chapman and Hall/CRC,
Boca Raton.

\item Nair, R., and Miller-Hooks, E. (2011). Fleet management for vehicle
sharing operations. \emph{Transportation Science }\textbf{45 }524--540.

\item Nair, R., Miller-Hooks, E., Hampshire, R.C., and Bu\v{s}i\'{c}, A.
(2013). Large-scale vehicle sharing systems: Analysis of V\'{e}lib'. \emph{%
International Journal of Sustainable Transportation }\textbf{7 }85--106.

\item Ramsay, J. O., and Silverman, B. W. (2005). \emph{Functional Data
Analysis. Second Edition}. Springer, New York.

\item Romano, E., Balzanella, A., and Verde, R. (2010). Clustering
spatio-functional data: A model based approach. In \emph{Classification as a
tool for research}. \emph{Studies in Classification, Data Analysis, and
Knowledge Organization}, pp.~167--175. Springer, Berlin, Heidelberg.

\item Ruppert, D. (2002). Selecting the number of knots for penalized
splines. \emph{Journal of Computational and Graphical Statistics} \textbf{11}
735--757.

\item Seber, G.A.F. (2004). \emph{Multivariate Observations.} Wiley, New
York.

\item Secchi, P. , Vantini, S. , and Vitelli, V. (2013). Bagging Voronoi
classifiers for clustering spatial functional data. \emph{International
Journal of Applied Earth Observation and Geoinformation} \textbf{22} 53--64.

\item Shaheen, S., Guzman, S., and Zhang, H. (2010). Bike sharing in Europe,
the Americas and Asia: Past, present and future. \emph{Transportation
Research Record: Journal of the Transportation Research Board }\textbf{2143}
159--167.

\item Shirota, S., and Gelfand, A.E. (2017). Space and circular time log
Gaussian Cox processes with application to crime event data. \emph{The
Annals of Applied Statistics }\textbf{11} 481--503.

\item Silverman, B.W. (1986). \emph{Density Estimation for Statistics and
Data Analysis.} Chapman and Hall/CRC, Boca Raton.

\item Streit, R.L. (2010). \emph{Poisson Point Processes: Imaging, Tracking,
and Sensing.} Springer, New York.

\item Vogel, P., Greiser, T., and Mattfeld, D.C. (2011). Understanding
bike-sharing systems using data mining: exploring activity patterns. \emph{%
Procedia Social and Behavioral Sciences }\textbf{20 }514--523.

\item Wu, S., M\"{u}ller, H.-G., and Zhang, Z. (2013). Functional data
analysis for point processes with rare events. \emph{Statistica Sinica }%
\textbf{23} 1--23.
\end{description}

\end{document}